# Quasi-three body systems - properties and scattering

M. Ya. Amusia[1, 2]

[1] *Racah Institute of Physics, the Hebrew University, Jerusalem 91904, Israel*
[2] *Ioffe Physical-Technical Institute, 194021 St. Petersburg, Russia*

**Abstract**

We investigate systems of three mutually interacting particles with masses $m_e$, $m_\mu$, M that obey the following inequality $m_e \ll m_\mu \ll M$. Then the three-body problem reduces to the two-body scattering or structure of $m_e$ in the field of the pseudo-nucleus $m_\mu M$. We calculate analytically the properties of considered systems, such as the scattering cross-sections, hyperfine splitting, Auger decay of exited states and Lamb shits, presenting them as expansions in powers of the parameter $\beta = m_e / m_\mu \ll 1$.

**1.** The aim of this presentation is to demonstrate that one can reduce a three-body system consisting of mutually interacting particles with masses $M$, $m_\mu$ and $m_\mu$ interrelated by the inequality $m_e \ll m_\mu \ll M$, to a much simpler two-body problem. Assuming without loss of generality that $M = \infty$, one can present the properties of $m_e m_\mu M$ system and the characteristics of the processes with participation as these particles as expansions in powers of small parameter $\beta = m_e / m_\mu \ll 1$. Along with integer powers of pure $\beta$ these expansions include also at least semi-integer powers and logarithms $|\beta \ln \beta| > \beta$. The account of this small parameter reduces considerably the solution of the considered three-body problem. We employ the atomic, $e = m_\mu = \hbar = 1$ system of units.

**2.** As an example, it is convenient to consider a system that consists of an electron $e$, $\mu^-$-meson and a proton $p$ for scattering or $e\mu^-\alpha$ ($\alpha \equiv {}^4He^{++}$ as well as hellion $h \equiv {}^3He^{++}$ ${}^3He^{++}$) for the structure studies. In this case, $\beta \approx 1/207$ Let us start with low-energy scattering. As a target, consider muonic hydrogen atom $H_\mu(\mu^- p)$ in its ground 1s state. In Hartree approximation one has [1]:

$$\sigma_{tot}^{1s} = 4\beta^2 \pi a_\mu^2. \tag{1}$$

that is, counterintuitively, much less than the geometric value $4\pi a_\mu^2$, where $a_\mu$ is Bohr's radius of $H_\mu$. Low energy $E$ means $E \ll 1/2$. The reason for the extra smallness $\beta^2$ in the cross section is the fact that in order to "feel" the distance $a_\mu \approx 1$ the incoming electron needs to have high energy $E \geq 1/\beta \approx 13.6/\beta^2$ eV for which the collision cross-section is small.

Surprising is the result for the polarization contribution to the scattering amplitude upon $H_\mu$

$$f_{pol}^{1s} = \frac{4\beta}{3\pi} \sum_n |\langle 1s | \vec{r}_\mu | n \rangle|^2 \int_0^\infty \frac{dk}{(k^2/2\beta) + \omega_{n1s}} = \frac{8\beta^{3/2}}{3\pi} \sum_n |\langle 1s | \vec{r}_\mu | n \rangle|^2 / (2\omega_{n1s})^{1/2} \approx 3\beta^{3/2} a_\mu. \tag{2}$$

Summation goes over all exited states of $H_\mu$, including integration that includes continuous spectrum. The main contribution comes from $k \geq \sqrt{2\omega_{n1s}} \approx 1$, where $\omega_{n1s} \equiv E_{1s}^\mu - E_n^\mu$.

The scattering of electron $e^-$ and positron $e^+$ differs by the projectile sign only, so the following expression presents the low-energy elastic scattering cross sections

$$\sigma_{tot}^{e\mp} = 4\beta^2 \pi a_\mu^2 \left(1 \pm 6\beta^{1/2} + 9\beta\right). \tag{3}$$

Even for rather small $\beta \approx 1/207$, one has cross-sections $\sigma_{tot}^{e-} \approx 1.46\beta^2 a_\mu^2$ and $\sigma_{tot}^{e+} \approx 0.62\beta^2 a_\mu^2$ that differ by a factor bigger than 2.

The expansion in $\beta$ one can apply also to targets that are bound states of a particle and an antiparticle, e.g. as $p\bar{p}, \mu\bar{\mu}$. In these cases the Hartree contribution disappears and the first non-vanishing term becomes the contribution of polarization that have an extra power in $\beta$: $\sigma_{tot}^{e\mp} = 36\beta^3 \pi a_\mu^2$.

**3.** Contrary to the ground state, the target polarization determines entirely the scattering upon an exited state $2s$ leading to the amplitude (2) with $1s$ substituted by $2s$. The contribution of the Hartree term is bigger than given by (1) due to doubling of $2s$ size as compared to $1s$. However, in this case the polarization contribution (2) dominates. The reason is the smallness of the energy of the closest to $2s_{1/2}$ excited level $2p_{1/2}$, $\omega_{2p1/2,2s1/2} \approx 0.82\alpha^3 \ll \omega_{2p1s} = 3/8$ where $\alpha \approx 1/137$ is the fine structure constant. Therefore, in the sum over $n$ in (2), it is sufficient to retain only one term with $n = 2p$. The relation for polarization contribution becomes for exited $H_\mu$

$$f_{pol}^{2s} = \frac{8\beta^{3/2}}{3\pi} \left|\langle 2s_{1/2} | \vec{r}_\mu | 2p_{1/2} \rangle\right|^2 / (2\omega_{2p1/2,2s1/2})^{1/2} \approx 25.3 (\beta/\alpha)^{3/2} a_\mu \approx 13.8 a_\mu. \tag{4}$$

This is by $\beta^{-1}$ bigger than the contribution of the Hartree term. Similar is the situation for electron scattering upon such targets as $p\bar{p}$, $\mu\bar{\mu}$ with $\beta_{\mu\bar{\mu}} = m_e / m_\mu^{red}$, where $m_\mu^{red} = m_\mu / 2$.

The radial dependence of polarization potential that corresponds to the amplitudes (3) and (4) is qualitatively different from the accepted expression that at $r \to \infty$ is of the form $U_{pol}(r) = -\alpha_\mu / 2r^4$, where $\alpha_\mu$ is the static dipole polarizability of $H_\mu$. Since in the denominator of (3) the projectile's virtual excitations, counterintuitively, are much less important, the target behaves as an object in the degenerate state with corresponding polarization potential $U_{pol}(r) = -\gamma_\mu / r^2 \sim -\beta/r^2$. This long-range behavior is correct up to $r_{max}^{1s} \approx 1/\sqrt{2\beta\omega_{2p1s}} \approx 1/\sqrt{3\beta/8}$ and $r_{max}^{2s1/2} \approx 1/\sqrt{2\beta\omega_{2p1/2,2s1/2}} \approx 1.1/\sqrt{\beta\alpha^3}$ for atom $H_\mu$ in the ground $1s$ and exited states $2s$, respectively.

The same argumentation is valid for Van-der-Waals potential between two atoms, A and B, e.g. normal hydrogen $H$ and $H_\mu$ or positronium $Ps$ and $H_\mu$. Its radial long-range dependence instead of $W^{AB} \sim -1/r^6$ becomes $W^{HH_\mu} \sim -1/r^4$.

The smallness of $\beta$ greatly simplifies the calculation of $Ps$ scattering upon $H_\mu$ since $H_\mu$ weakly perturbs the relative motion of $e^-$ and $e^+$ in $Ps$ that remains purely hydrogenlike.

4. The only open channel for inelastic electron scattering at $E<1/2$ is bremsstrahlung, BS [2]. Its cross-section $\omega d\sigma_{BS}^{e^{\mp}}/d\omega$ has an ordinary contribution that is proportional to the elastic scattering cross-section and a contribution that represent radiation of the target $H_\mu$, polarized by the projectile $e^{\mp}$. Of interest is the expression for the ratio $\eta_\mu^{\mp}$ of $\omega d\sigma^{BS}/d\omega$ and $\sigma_{tot}^{\mp}$:

$$\eta_\mu^{\mp}(E,\omega_\gamma) = \frac{\omega_\gamma d\sigma^{BS}/d\omega_\gamma}{\sigma_{el}} = \frac{8}{3\pi}\alpha^3\beta^{-2}E\left[1\mp\beta\frac{9\omega_\gamma^2}{2E}+\left(9\beta\frac{\omega_\gamma^2}{E}\right)^2\ln\left(\frac{E}{\omega_\gamma}\right)\right]. \tag{5}$$

Here $\omega_\gamma$ is the energy of the emitted photon. If $E$ is not very small, $E\leq 1$, thus being close to ionization potential of $H_\mu$, and $\omega_\gamma\ll 1$, the ratio $\eta_\mu^{\mp}$ is bigger than that for ordinary hydrogen by the factor $\beta^{-2}\gg 1$. This ratio increases dramatically when $E$ approaches $\omega_{1s,2p}$, where in the BS cross-section the radiation of the polarized target is bigger than the ordinary contribution by orders of magnitude. The ratio $\eta_\mu^{\mp}(E,\omega)$ reaches its maximum value at the resonance $\omega_{1s,2s}$. To obtain the ratio there, one has to take into account very small width of the resonance $\Gamma_{1s2p}$. As a result, the ratio $\eta_\mu^{\mp}(E\geq\omega_{1s2p},\omega_{1s2p})\sim\beta^{-3}\approx 10^7$.

5. Along with scattering processes, the smallness of $\beta$ simplifies considerably the calculation of such ground state characteristic as hyperfine splitting $\Delta^{HFS}$. Here a proper example is $\Delta^{HFS}$ of $1s_{1/2}^e 1s_{1/2}^\mu$ and $1s_{1/2}^e 2s_{1/2}^\mu$ states of $e^-\mu^-\alpha(h)$ atoms. The specifics of these objects as compared to ordinary hydrogen is much bigger size and polarizability of pseudo-nuclei $1s_{1/2}^\mu\alpha(h)$ and $2s_{1/2}^\mu\alpha(h)$.

Using small parameter $\beta$ one can present $\Delta^{HFS}$ for $1s_{1/2}^e 1s_{1/2}^\mu$ of the atom $e^-\mu^-\alpha$ with the so-called logarithmic accuracy in the following way [3]

$$\Delta_{HFS}^{e^-\mu^-\alpha} = \Delta_0[1-1.5\beta+1.96\beta^{3/2}-2\beta^2\ln\beta^{-1}]. \tag{6}$$

Here $\Delta_0 = 32\alpha^2\beta^2/3$ is the hyperfine splitting for $e$ and a unit positive charge without inner structure magnetic moment of a muon. It is remarkable that (6) gives $\Delta_{HFS}^{e^-\mu^-\alpha} = 4462.9$ MHz that is close to the experimental value 4464.95(6) MHz.

Similar to (6) but with different coefficients is the expression $\Delta^{HFS}$ for the excited $2s_{1/2}$ muon state [4]:

$$\Delta_{HFS}^{e^-\mu_{2s}^-\alpha} = \Delta_0[1-12.42\beta+4.32\beta^{3/2}-48.5\beta^2\ln\beta^{-1}]. \tag{7}$$

The numeric value of $\Delta_{HFS}^{e^-\mu_{2s}^-\alpha}$ is 4291.5 MHz. For $h$ we, naturally, come to a more complicated physical picture for the HFS than for the $\alpha$-particle. Indeed, here the nuclear spin $\mathbf{I}$ couples with the muon spin $\mathbf{s}_\mu$ that result in the momentum $\mathbf{F}_1 = \mathbf{I}+\mathbf{s}_\mu$. The latter couples with the electron spin $\mathbf{s}_e$ and finally gives the total moment of the whole atom $\mathbf{F} = \mathbf{F}_1+\mathbf{s}_e$. The investigation leads to an expression similar to (7), but with other coefficients resulting in numeric value of the hyperfine splitting 4056.24 MHz.

6. The excited states $1s^e_{1/2}ns^\mu_{1/2}$ of $e^-\mu^-\alpha(h)$ atoms are unstable. They can decay not only by emitting two photons but also via Auger-decay emitting the electron from the 1s-level. The big difference between electron $a_e \sim 1/\beta$ and muon $a_\mu \approx 1$ orbits permits easily to calculate the first non-vanishing term and the lowest order correction to it in powers of $\beta$ [5]:

$$W^{ms^\mu_{1/2},ns^\mu_{1/2}}_{Aug} \approx \frac{4k_{mn}}{9}\left(1-\frac{\pi}{k_{mn}}\right)\beta^4 \left|\int_0^\infty dr \varphi_{ms}(r)r^4\varphi_{ns}(r)\right|^2 \Rightarrow 3.95\beta^4 k_{21}\left(1-\beta^{1/2}\frac{\pi}{k_{21}}\right)\Bigg|_{m=2,n=1}. \quad (8)$$

Here $\varphi_{ns}(r)$ are pure hydrogenlike wave functions for $Z=2$ nuclear charge and $k_{nm}=\sqrt{2(\varepsilon_m-\varepsilon_n)-\beta/2}$ is the linear momentum of the emitted electron. The expression (8) permits to estimate the ration of probabilities of two-photon $W_{2\gamma}$ and Auger $W_{Aug}$ decay rates as $W_{2\gamma}/W_{Aug} \sim \beta^{3/2} \approx \alpha^{3/2} \ll 1$. The expression (8) is accurate enough only for sufficiently low excited muon states, for which $n^2\beta \ll 1$ or $n \ll 14$. If this condition non-valid, electron $e$ essentially perturbs the $n_\mu$-state.

Numerically, the Auger-decay lifetime of $e^-\mu^-\alpha(h)$ in the $1s^e_{1/2}2s^\mu_{1/2}$ state is 8.8x10$^{-8}$ sec while muon lifetime is 2.2x10$^{-6}$ sec. Therefore, muon is stable enough for experimentation with the $1s^e_{1/2}2s^\mu_{1/2}$ state of exotic atoms $e^-\mu^-\alpha(h)$.

7. Of interest is to study the Lamb shift of electronic states in $e^-\mu^-\alpha(h)$. Here again the well-defined hierarchical structure and small affect upon the intermediate particle by the outer one opens up a possibility to perform quite accurate calculations analytically, presenting the results as main hydrogenlike term along with corrections, presented as expansions in $\alpha$ (radiative corrections) and $\beta$. Our theoretical predictions [6] for the following transitions in the atom $1s^e_{1/2}1s^\mu_{1/2}h$ are $f_{1s2s}=2467046(3)$ GHz, $f_{2s2p1/2}=9.3(4)$ and $f_{2s2p3/2}=20.3(4)$ GHz, while for $1s^e_{1/2}1s^\mu_{1/2}\alpha$ it is $f_{1s2s}=2467149(4)$ GHz, $f_{2s2p1/2}=9.1(6)$ and $f_{2s2p3/2}=20.1(6)$ GHz. The cases of exited atoms $1s^e_{1/2}2s^\mu_{1/2}\alpha(h)$ are not yet considered, but their lifetime is in principle enough for measuring the Lamb shift. The polarizability of the pseudo-nucleus $2s^\mu_{1/2}\alpha(h)$ is much bigger than that of $1s^\mu_{1/2}\alpha(h)$.

8. A number of other systems can be treated using the presented here approach. The inter-particle interaction can be non-Coulombic. One can study e.g. scattering and binding electrons and hadron targets as consisting of quarks, or quark and antiquarks, with the parameter $\beta \ll 1$. It is essential that such an approach help correctly and relatively simple to take into account the action of targets' polarizability on the electron (positron) scattering. This same approach can be of use in studying the difference in the proton radius, obtained from $H$ and $H_\mu$ spectra [7].